\documentstyle [12pt,epsf]{article}

\newcommand{\postscript}[2]
 {\setlength{\epsfxsize}{#2\vsize}
\setlength{\epsfysize}{#2\vsize}
  \centerline{\epsfbox{#1}}}

\newcommand{\postscriptfx}[1]
 {\setlength{\epsfxsize}{6in}
\setlength{\epsfysize}{6in}
  \centerline{\epsfbox{#1}}}

\begin{document}
\baselineskip 7.5 mm

\def\thefootnote{\fnsymbol{footnote}}
\baselineskip 7.5 mm

\begin{flushright}
\begin{tabular}{l}
UPR-677-T \\
NSF-ITP-96-09 \\
February, 1996 
\end{tabular}
\end{flushright}

\vspace{2mm}

\begin{center}

{\Large \bf Phase Transitions and Vacuum Tunneling Into Charge and Color
Breaking Minima in the MSSM.}
\\ 
\vspace{8mm}

\setcounter{footnote}{0}

Alexander Kusenko\footnote{ email address:
sasha@langacker.hep.upenn.edu; address after October 1, 1996:
Theory Division, CERN, CH-1211 Geneva 23, Switzerland
}$^{1,2}$,  
Paul Langacker\footnote{ email address: pgl@langacker.hep.upenn.edu}$^{1,2}$
\\ and \\
Gino Segre\footnote{email address: segre@dept.physics.upenn.edu}$^{1}$ 
\\
$^1$Department of Physics and Astronomy, University of Pennsylvania \\ 
Philadelphia, PA 19104-6396 \\
$^2$Institute for Theoretical Physics, University of California, Santa
Barbara, CA 93106 \\

\vspace{12mm}

{\bf Abstract}
\end{center}

The scalar potential of the MSSM may have local and global minima
characterized by non-zero expectation values of charged and colored bosons.
Even if the true vacuum is not color and charge conserving, the early
Universe is likely to occupy the minimum of the potential in which only
the neutral Higgs fields have non-zero vev's.  The stability of this false 
vacuum with respect to quantum tunneling imposes important constraints on
the values of the MSSM parameters.  We analyze these constraints using some
novel methods for calculating the false vacuum decay rate.  Some regions
of the MSSM parameter space are ruled out because the lifetime of the
corresponding physically acceptable false vacuum is small in comparison to
the present age of the Universe.  However, there is a significant fraction
of the parameter space that is consistent with the hypothesis that the
Universe rests in the false vacuum that is stable on a cosmological time
scale.   

\vfill

\pagestyle{empty}

\pagebreak

\pagestyle{plain}
\pagenumbering{arabic}
\renewcommand{\thefootnote}{\arabic{footnote}}
\setcounter{footnote}{0}

\pagestyle{plain}

\section{Introduction}

In the Standard Model color and electric charge are automatically
conserved because the only fundamental scalar field
is the Higgs boson, a colorless electroweak doublet.  The
Higgs potential has a continuum of degenerate minima, but these are all
physically equivalent, and without loss of generality one can always define
the unbroken $U(1)$ generator to be the electric charge.  
This is not the case in the Minimal Supersymmetric
Standard Model (MSSM), which employs a pair of Higgs doublets as well as 
a number of other scalar fields, the supersymmetric partners of quarks and
leptons.  Although the relative alignment of the two Higgs doublets' vacuum
expectation values in group space is physical, the minimum of the Higgs
potential (at least at tree level) preserves electric charge\footnote{This
need not be the case in a general model with two Higgs doublets.} as long as
the squark and slepton fields have vanishing classical values (see, {\it
e.\,g.}, Ref. \cite{hhg} and references therein).  However,
the full scalar potential of the MSSM may have additional charge and/or
color breaking (CCB) minima due to the vacuum expectation values of
charged and/or colored scalars. 

The existence of the CCB minima in the MSSM in addition to the 
acceptable standard-model-like (SML) minimum may have important 
physical consequences.
One might expect that the regions of parameter space for which there is a
global CCB minimum could be automatically excluded, thereby further
restricting theoretical predictions for the MSSM spectrum. However, one
must be careful in drawing such conclusions.  Just as the
cup being the lowest point on the golf course by no means guarantees that
the ball will end up there after being struck, the Universe at present may 
not be in its lowest possible energy state. 
Instead, it may rest in a false vacuum whose lifetime is large on a 
cosmological time scale.  The fundamental reason that makes
this possible is that quantum tunneling, a non-perturbative effect
responsible for the first-order phase transitions in field theory, naturally
introduces a time scale that is exponentially larger than the typical
scale that characterizes the effective potential.  Consequently, the
relaxation to the lowest energy state from some excited state may take
a very long time.  In particular, parameters
for which the local SML ``false vacuum'' has a lifetime large in comparison
to the age of the Universe may be acceptable, provided of course that the SML
minimum was populated first in the evolution of the Universe.  

The existence of local CCB
minima which were populated temporarily during the early stages in the
evolution of the Universe would also have dramatic implications for
cosmology and astrophysics. In particular, since baryon and lepton numbers
are spontaneously violated in the CCB vacua, their existence might have
important consequences for baryogenesis.  

Previous attempts \cite{early}-\cite{clm} to elucidate the structure of the
CCB minima in the MSSM met with serious difficulties. Some analyses
\cite{early,ghs,k,clm} attempted to find analytic constraints on CCB
minima.  However, such conditions are generally neither necessary nor
sufficient \cite{ghs} except for overly simplified toy models which
resemble the MSSM in some features, but cannot be used to draw firm
conclusions about the MSSM.  For this reason, recent studies
\cite{lp,b,clm} have employed extensive numerical analyses.  Secondly, the 
determination of whether a global minimum is ``dangerous'', or not,
must rely on a trustworthy calculation of the tunneling rates at present
and in the early Universe.  There is no reason why the Universe cannot
be resting in a false vacuum which has a very long (on the cosmological
scale) lifetime.  We therefore disagree with the restrictions imposed by
a number of authors \cite{k,b,clm} on the allowed MSSM parameter space,
which did not consider the corresponding tunneling rates.  The calculation
of the transition probability is more or less straightforward in the case
of a single scalar but becomes extremely difficult for a potential that
depends on several fields. Below, we address these difficulties and employ
a new technique \cite{ak1} to determine the lifetime of the false vacuum in
the case of the MSSM.   

We will see, in fact, that the SML minimum is effectively stable with
respect to the transitions to the corresponding CCB minima for a
substantial part of the allowed parameter space in the MSSM.  We will also
argue that,  due to the specific nature of the CCB minima, they would not
have been populated during the early stages of the evolution of the
Universe, except for some small regions of parameters.  On the other 
hand, the stability of the color and charge conserving vacuum on the time
scales of order of the age of the Universe imposes important constraints on 
theoretical models and can provide guidance for future experimental
searches. 

The paper is organized as follows.  In Section 2, we discuss
general features of the SML vacuum decay in the MSSM. 
In Section 3, we consider the electroweak phase transition in the 
early Universe, in particular the issue of the $SU(3)\times SU(2)\times
U(1)$-symmetric vacuum stability before the transition. 
In sections 4 and 5, we study the zero-temperature tunneling rates
numerically.  The method used to compute the transition probability
is described in Appendix A, while Appendix B contains an approximate 
description of tunneling in the limit of a very deep true minimum. 

\section{Essential aspects of the MSSM vacuum stability}

We begin by considering a simplified version of the MSSM potential. 
As was emphasized in \cite{chh,lp}, the third generation requires the most
attention in connection with the issue of color and charge breaking, because
the  the CCB minima associated with the the large Yukawa coupling are the
most dangerous\footnote{This statement was disputed in \cite{clm}, where it
was argued that in the limit of small Yukawa coupling both the cubic and
the quartic terms are small and that only their relative values 
affect the depth of the CCB minimum.  This is true, as long as one
ignores the quartic D-terms which, in fact, become more important in the
small Yukawa coupling limit.  However, as was shown in Ref. \cite{ehnt},
the height of the barrier separating the SML minimum from the CCB minimum is
roughly proportional to $1/y_{min}^2$, where $y_{min}$ is the smallest Yukawa
coupling associated with the fields that acquire non-zero vev in the CCB
minimum.  The corresponding tunneling rates are greatly suppressed for
small $y$.  This point is illustrated in the toy model discussed below (see
also Ref. \cite{chh}).  Thus the CCB minima in which only the 
squarks of the third generation have non-zero vev play the most
important role in our considerations.}.  We will see shortly that the
tunneling rate into a CCB minimum is roughly proportional to $\exp
(-c/y^2)$, where $y$ is the corresponding Yukawa coupling and $c$ is a
constant.   

We begin by considering a model defined by the superpotential:

\begin{equation}
W=y t_{_L} t_{_R} H_2 + \mu H_1 H_2,
\end{equation}
where $t_{_L}$ and $t_{_R}$  denote the top quark superfields, and the
$H_1$ and $H_2$ are the MSSM Higgs bosons.  At this point we ignore the
leptons, lighter quarks, and the electrically charged Higgs components.  
The resulting scalar potential, including the soft SUSY breaking terms,
is, at tree level,   

\begin{equation}
V=V_{2}+V_3+V_4
\label{V}
\end{equation}
where 

\begin{eqnarray}
V_2 & = & m_1^2 H_1^{0^2} + m_2^2 H_2^{0^2} +2 m_3^2 H_1^0 H_2^0 
+ m_{\tilde{t}_{_L}}^2 \tilde{t}_{_L}^2 + m_{\tilde{t}_{_R}}^2
\tilde{t}_{_R}^2  \label{V2} \\ 
V_3 & = & -2 A H_2^0 \tilde{t}_{_L} \tilde{t}_{_R} - 2 \mu H_1^0
\tilde{t}_{_L} \tilde{t}_{_R}  \label{V3} \\ 
V_4 & = & \tilde{t}_{_L}^2 \tilde{t}_{_R}^2 + \tilde{t}_{_L}^2 H_2^{0^2} +
\tilde{t}_{_R}^2 H_2^{0^2} + V_{_D} 
\label{V4}  
\end{eqnarray}

For the $ SU(3) \times SU(2) \times U(1)$ gauge group, the $D$-terms are 

\begin{equation}
V_{_D}= \frac{1}{8y^2} \ \left [ 
g_1^2(H_1^{0^2}-H_2^{0^2}-\frac{1}{3}\tilde{t}_{_L}^2
+\frac{4}{3} \tilde{t}_{_R}^2)^2 + g_2^2
(H_1^{0^2}-H_2^{0^2}+\tilde{t}_{_L}^2)^2+  
\frac{4}{3}g_3^2 (\tilde{t}_{_L}^2-\tilde{t}_{_R}^2)^2 \right ]
\label{VD}
\end{equation}
Here the color indices are suppressed. We have absorbed the Yukawa coupling
in equations (\ref{V}-\ref{VD}) by the 
redefinition of the fields $\phi \rightarrow \phi/y$ and of the scalar
potential $V \rightarrow y^2 V$.  Also, all the fields are made real by a 
rephasing, and the complex phases are absorbed into the definitions of $A$
and $\mu$ parameters.  (There are strong experimental limits on such
phases, which force $A$ and $\mu$ to be nearly real; see Ref. \cite{dgh} for
reviews of these constraints.) 

In the $\tilde{t}_{_L}=\tilde{t}_{_R}=0$ hyperplane, equation (\ref{V})
describes the usual MSSM Higgs potential.  The constraint 

\begin{equation}
m_1^2 m_2^2 < (m_{3}^{2})^2 < \left ( \frac{m_1^2 +m_2^2}{2} \right )^2 
\label{hptn}
\end{equation}
ensures the existence (first inequality) and stability (second inequality)
of the minimum with the correct pattern of electroweak symmetry
breaking.  The latter inequality results from requiring that the quadratic
term is positive definite along the flat directions of the quartic term
$V_4$.  

Much of our discussion will concentrate on the effects of the trilinear 
terms $V_3$.   If $A$ and/or $\mu$ are large enough, the potential acquires 
an additional local, or global, minimum at some point outside the
$\tilde{t}_{_L}=\tilde{t}_{_R}=0$ hyperplane.  In this case, the
electromagnetic $U(1)$, color  
$SU(3)$, as well as some other symmetries ({\it e.\,g.}, the global
$U(1)_{baryon}$) will be spontaneously broken by the non-zero vev of
$\tilde{t}_{_L}$ 
and $\tilde{t}_{_R}$.  For example, the potential (\ref{V}) has, for
appropriate 
values of $A$ and $\mu$, four degenerate CCB minima mapped onto each other
by the following reflections:

\begin{equation}
S_1=\left \{ \begin{array}{c}
             H_1 \rightarrow - H_1 \\
             H_2 \rightarrow - H_2 \\
             \tilde{t}_{_L} \rightarrow - \tilde{t}_{_L} 
            \end{array} \right \}; \ \ 
S_2=\left \{ \begin{array}{c}
             H_1 \rightarrow - H_1 \\
             H_2 \rightarrow - H_2 \\
             \tilde{t}_{_R} \rightarrow - \tilde{t}_{_R} 
            \end{array} \right \}; \ \ 
S_3=S_1 S_2 =\left \{ \begin{array}{c}
             \tilde{t}_{_L} \rightarrow - \tilde{t}_{_L} \\
             \tilde{t}_{_R} \rightarrow - \tilde{t}_{_R} \\
            \end{array} \right \} \ \ 
\end{equation}

The gauge $SU(3)\times SU(2)\times U(1)$ symmetry is broken by the 
non-zero $\tilde{t}_{_L},\ \tilde{t}_{_R}$ and $H_2$ vacuum expectation
values down to an  $SU(2)$ subgroup of the color $SU(3)$.  

Evidently, for some otherwise reasonable values of the parameters, the
potential may have a global CCB minimum.  If this is the case, we would
like to estimate the tunneling probability from the SML to the CCB
minimum. 

The semi-classical calculation of the false vacuum decay width was done in
Refs. \cite{vko,c,cc} for the case of a single scalar field, $\phi(x)$.  
For a recent review of tunneling we refer the reader to Ref. \cite{v}. 
The corresponding path integral is dominated by the field configuration
$\bar{\phi}(x)$ called the ``bounce'' and can be evaluated using the saddle
point method  \cite{c,cc}.
The bounce, being the stationary point of the Euclidean
action, is the non-trivial solution of the corresponding Euler-Lagrange
equation obeying certain boundary conditions.
 
The transition probability per unit volume in the semi-classical limit
\cite{cc} is
 
\begin{equation}
\Gamma/{\sf V} = {\sf A} e^{-S[\bar{\phi}]/\hbar}, 
\label{1}
\end{equation} 
where $S[\bar{\phi}]$ is the Euclidean action of the bounce, a classical
solution to the variational equation $\delta S=0$.  

For the Universe to have decayed to the global minimum, the transition
has to take place within a four-volume of size, roughly, $t_0^4$, where
$t_0 \sim 10^{10}$ years is the age of the Universe.  Taking the
pre-exponential factor in (\ref{1}) to be of order 
$(100 \ {\rm GeV})^4$, one obtains $(\Gamma/{\sf V}) t_0^4 \sim 1$ for
$S_{_E}[\bar{\phi}]/\hbar \sim 400$.  Therefore, a false vacuum whose decay
rate is characterized by $S_{_E}[\bar{\phi}]/\hbar >400$ can safely be
considered stable. 

The presence of many scalar fields in the potential introduces a number of
complications which will be addressed below\footnote{See also
Refs. \cite{ak1,ak2,klw}.}.  However, our immediate goal 
is to obtain a crude estimate of the false vacuum decay rate.  We therefore
make a further simplification and reduce the tunneling problem to 
that of a one-component case. 

Suppose the energy difference between the two minima $\Delta
V= \epsilon^4 $ is small in comparison to the height of the barrier. 
Then the thin-wall approximation is appropriate and
the Euclidean action of the bounce of size $R$ is given by 

\begin{equation}
y^2 S_{_E}  = -2 \pi^2 \frac{R^4}{4} \Delta V+ 2 \pi^2 R^3
\int_{\phi_{_{SML}}}^{\phi_{_{CCB}}}  
\sqrt{2V(H)} \ d \phi
\end{equation}
where the appearance of the Yukawa coupling on the left-hand side of the
equation is due to the fact that the fields have been scaled by a factor
$y$ in the beginning. 

The bounce $\bar{\phi}(x)$ corresponds to the extremum of the Euclidean action
with respect to $R$, which is reached for the critical size of the
``bubble''  $R_c=3 S_1/ \epsilon^4 $, where $S_1=  \int \sqrt{2V(H)} \ d
\phi$. The corresponding action is 

\begin{equation}
S_{_E}[\bar{\phi}] \approx \frac{27\pi^2}{2 y^2} \frac{S_1^4}{\epsilon^{12}}
\label{S_thinwall}
\end{equation} 

Some comments are in order.  First, we observe that, as was asserted above,
the tunneling rate is very sensitive to the value of the Yukawa coupling.
Therefore, the minima associated with the third generation of squarks are
the most interesting.  Second, it is well-known that
the thin-wall approximation works well only for very small values of
$\epsilon$.  On the other hand, $S_{_E}[\bar{\phi}] < 400$ in
(\ref{S_thinwall}) corresponds to $\epsilon/S_1^{1/3} > 
[(27\pi^2/2 y^2)/400]^{1/12} \approx 0.9$, which is not the thin wall regime.
This means that whenever the thin-wall limit is
a good approximation, the transition we are interested in will not take
place on the relevant time scale.  The ``dangerous'' CCB minima lie, as
a rule, outside the domain of validity of the thin-wall approximation. 

In Appendix B, we find an approximate representation of the bounce in the 
opposite limit, which we call a ``thick-wall approximation''. Unfortunately,
the phenomenologically acceptable values of the trilinear term can be
approximated by neither thin-wall,  nor thick-wall  limiting expressions.
For this reason, one must resort to a numerical analysis to determine the
fate of the false SML vacuum in the presence of the lower lying CCB vacua.

We will focus on the CCB minima associated with the scale of order the 
electroweak and the SUSY breaking scales, in contrast to other studies 
\cite{fors,rr} that discussed the possibility of a potential CCB minimum
characterized by a planckian, or a GUT-scale scale vev.  

It was argued in Ref. \cite{fors} that, if the MSSM is to be
incorporated in some Grand Unified Theory (GUT) characterized by the scale 
$M_{_G}$, some new constraints should be imposed on the MSSM parameters to
eliminate the possibility of a global CCB minimum developing at some scale
$Q$, $1 \ {\rm TeV} \ll Q < M_{_G}$.  We note that symmetry
restoration at a large energy scale is not required for the self-consistency
of a spontaneously broken gauge theory.  The latter is characterized by 
a symmetric action and an asymmetric vacuum.  Cosmological data may, at least 
in principle, provide a test of whether the symmetric ground state existed
in the Early Universe.  However, the finite-temperature field theory
describing an expanding universe has a different effective potential from
that of the $T=0$ case.  Most of the flat directions of the
tree-level potential are lifted by terms of order SUSY breaking scale
\footnote{In addition to finite-temperature and tree-level breaking of
supersymmetry, there are SUSY breaking terms associated with the
metastability of the false vacuum \cite{ak3}.}. 
Therefore, the CCB minima of the kind dealt with in Ref. \cite{fors} may
not be present in the Early Universe. 
Furthermore, as was pointed out in Ref. \cite{rr}, the existence of a 
global CCB minimum at such a large value of the vev is irrelevant 
for the low-energy physics because the color and charge conserving vacuum
is effectively stable with respect to tunneling into the CCB minimum of
that kind.  

This is an example of a phenomenon which is similar
to the usual (perturbative) decoupling.  Tunneling into a ``very deep'' 
vacuum occurs at a rate which is independent of the depth of such a minimum
and is determined only by the magnitude of the field and the steepness of
the potential at some well defined point, the ``escape point'', $\phi_e$
({\it c.\,f.} the discussion in our Appendix B).  The shape of the
effective potential at the scale $Q \gg \phi_e$ has no effect on the
transition probability, and therefore the low-energy physics is
independent of the physics at  $Q \gg \phi_e$.  This is analogous to 
decoupling in perturbation theory, even though the perturbative decoupling
theorems may not apply to non-perturbative effects such as tunneling. 

In practice, this decoupling allows us to treat the ultra-deep minima on
the same footing as the CCB minima of the tree-level potential.  Also,
since all the relevant dimensionful quantities, including $\phi_e$, are of
order $100$ GeV to $1$ TeV, the radiative corrections to the effective
potential have a very small effect on the tunneling rates.  Therefore, it
is well-justified to use a tree-level potential for evaluating the
stability of the SML vacuum and determining the allowed regions of
parameters in the MSSM.  In a future study, we plan to further refine our
present results by taking into account the radiative corrections to the
effective potential.   

Closely related is the issue of the directions in the scalar sector of the
MSSM along which the tree-level potential appears to be unbounded from
below (UFB).  These directions are chosen to zero the quartic terms in the
tree-level potential. Usually, the one-loop radiative corrections rectify
the situation by introducing positive definite quartic terms of the
type $ STr M^4 \log(M^2/Q^2)$.  If this is the case, the full effective 
potential  turns out to have only a very deep CCB minimum and 
is not unbounded from below.  The latter may be
separated from any other vacuum by a high enough barrier to make the
presence of such a CCB minimum irrelevant.  
To determine whether a certain region of the parameter space must be
excluded,  one must again examine the corresponding tunneling rates.  
However, since in any case the tunneling rate is determined by the shape of
the potential at the vev's of order a few TeV, one can treat the UFB
directions as if they were leading to a very deep minimum.  We stress that
this is true, regardless of whether the given direction is a UFB direction
of the exact effective potential, or only of its finite-order
(tree-level, one-loop, {\it etc.}) approximation.

\section{CCB minima in the early Universe} 

If the effective potential has more than one minimum, then the determination
of the physical vacuum requires consideration of the evolution of the
Universe.  For the MSSM, the analysis is complicated by the
presence of flat directions (see, {\it e.\,g.}, Ref. \cite{gkm}), lifted
only by terms of order the SUSY breaking parameters, along which the scalar
field may acquire a large vev and fall out of thermal equilibrium in the
early Universe.  We leave this question to future study and assume that 
the electroweak phase transition proceeds in the usual manner, from an 
$SU(3)\times SU(2)\times U(1)$-symmetric to a broken phase.  

This assumption is well justified for the types of CCB minima we consider. 
It is well-known that in inflationary models the large fluctuations of the
scalar fields may populate some color and charge breaking minimuma.  This is
true of both local and global minima.  If the vacuum expectation value of
the scalar field in that minimum is large in comparison to the reheating
temperature, the Universe may "freeze" in that minimum.  However, the CCB
minima we consider, unlike those of Refs. \cite{fors,rr}, have vev's of
the order of the electroweak scale and usually disappear at temperatures 
of order 1~TeV, except for the models with non-trivial (and atypical) 
symmetry restoration pattern like that of Ref. \cite{lp}. 
Since most inflationary models predict much higher reheating temperatures, 
one can assume that at $T \sim 1$~TeV the $SU(3)\times SU(2)\times
U(1)$ is unbroken. 

The main effect of the temperature-dependent corrections to the effective
potential is contributions of order $T^2$ to the quadratic terms.  
The trilinear terms also receive some corrections, linear in $T$.  

The depth of the CCB minima depend on the relative values of the squark mass
terms ($m_0$) on one hand, and $A$ or $\mu$ on the other.  At finite
temperature,  positive mass-squared terms proportional to $T^2$ appear
in the effective potential and lead, at some critical temperature, 
$T_c \sim 100$ GeV, to the disappearance of the SML minimum.  Since $A, \
\mu$ and  $m_0$ are allowed to be large in comparison to $T_c$, it is
possible that as the Universe cooled the negative energy CCB minima formed
at some $T_{_{CCB}}>T_c$, before the Universe was cold enough to undergo
the electroweak phase transition to the SML minimum.  This would allow for
the possibility of a transition from an $SU(3)\times SU(2)\times
U(1)$-symmetric to the color and charge breaking phase.  One then must
consider whether the transition to the CCB minimum actually occurred, and,
if so, what happened subsequently. 

Then, {\it a priori} there are three 
possibilities: (i) the Universe may stay in the symmetric minimum until 
the temperature reaches $T_c$ and the usual electroweak transition takes
place; (ii) the Universe may go to the CCB minimum and freeze there; 
and (iii) the transition from the symmetric to the SML minimum may occur in
two stages: first the transition from the unbroken to the CCB phase takes
place and only then the Universe may tunnel to the SML vacuum.  Clearly,
the second possibility is excluded empirically.  

Possibility (iii) is intriguing.  The idea of a
multi-stage phase transition, in the course of which the gauge group 
might change a number of times finally arriving at the Standard Model group,
is not new.  An evolution of this kind could have important
implications for magnetic monopoles \cite{lpi}, charge asymmetry of the
Universe and magnetic field generation \cite{d}, baryogenesis \cite{ad}, 
{\it etc.}  The sufficient (though not necessary) conditions for the first
part of this scenario, the transition from the symmetric to the CCB phase,
will be derived below.  However, we have not found a model in which a 
second-order phase transition from the CCB minimum to the SML minimum would
subsequently take place.  It appears to be a generic feature that even when
the CCB minima can be populated via a second-order phase transition, they
are separated from the SML minimum by a thick barrier.  
If the Universe is stuck in the false CCB vacuum (which, in this scenario,
must have a higher vacuum energy than the SML phase), this would trigger 
a new stage of inflation which could only be ended by a first order phase
transition characterized by a relatively low tunneling rate.  It was shown
in Ref. \cite{hgw} that this sort of transition 
proceeds via bubble nucleation at a rate which is too slow to catch up
with the expanding volume.  The bubbles of the true vacuum would neither 
collide, nor percolate, preventing the Universe from reheating.  Therefore, 
the two-stage electroweak phase transition could not have taken place in
the early Universe. 

Let us consider the possibility that a CCB minimum exists at some
temperature $T>T_c$. At that temperature, the mass matrix of the third
generation squarks (ignoring the rest of the squarks and sleptons) 
in the $SU(2)\times U(1)$-symmetric phase ($H_1=H_2=0$) is of the form:

\begin{equation}
\left ( \begin{array}{cc}
        m^2_{\tilde{t}_{_L}}+c_{_L} T^2 & 0 \\
        0 & m^2_{\tilde{t}_{_R}}+c_{_R} T^2 
        \end{array}  \right )
\label{MT}
\end{equation}

Suppose that it has at least one negative eigenvalue, which makes the 
$SU(2)\times U(1)$-symmetric vacuum unstable.  Since $T>T_c$, the SML
minimum does not yet exist.   The decay of the unstable vacuum will result
in the creation of the CCB condensate by a second-order phase transition.
At the same time, we assume that all the necessary conditions
have been applied to constrain the masses-squared  of squarks to be
positive (and large enough) in the SML minimum at zero temperature, where
the mass matrix is  

\begin{equation}
\left ( \begin{array}{cc}
        m^2_{\tilde{t}_{_L}}+m_t^2 & A H_2+\mu H_1 \\
        A H_2+\mu H_1 & m^2_{\tilde{t}_{_R}}+m_t^2 
        \end{array}  \right ),
\label{M0}
\end{equation}
where the $m_t$ is the top quark mass. 

\begin{figure}
\postscript{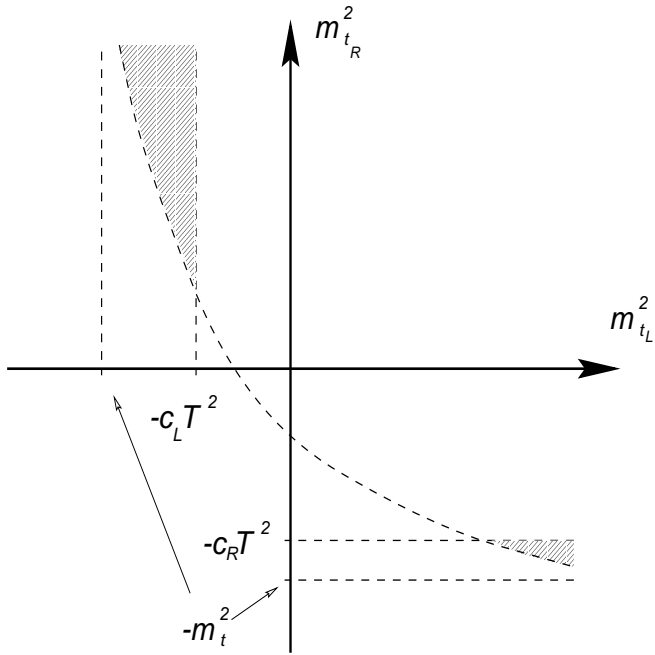}{0.4}
\caption{Region of parameters which can be ruled out by requiring
stability of the {$SU(2)\times U(1)$} symmetric minimum above the
electroweak transition temperature.}
\label{fig1}
\end{figure}

We would like to see whether there is a region of parameter space in which
the matrix (\ref{M0}) has only positive eigenvalues, while the matrix
(\ref{MT}) has a negative eigenvalue.  It is 
easy to see that this is only possible if either $c_{_L} T_c^2<m_t^2$, or
$c_{_R} T_c^2<m_t^2$. The shaded region shown in Figure \ref{fig1} 
corresponds to the additional domain of parameters which can be ruled out
by requiring stability of the $SU(2)\times U(1)$ symmetric minimum above
the electroweak transition temperature.  This domain comprises two regions
in which $m_{\tilde{t}_{_L}}^2$ and  $m_{\tilde{t}_{_R}}^2$ differ in sign
(Figure \ref{fig1}). 
The hyperbola $(m^2_{\tilde{t}_{_L}}+m_t^2)(m^2_{\tilde{t}_{_R}}+m_t^2)=(A
H_2+\mu H_1)^2$, where the values of $H_1$ and $H_2$ are computed at the
SML minimum, outlines the 
domain of positive determinant of the matrix in equation (\ref{M0}). 
If one requires that the squark masses be greater than 45 GeV, in
accordance with the current experimental limits, it would further reduce
the area of the corresponding (shaded) regions in Figure \ref{fig1}.  

The coefficients $c_{_L}$ and $c_{R}$ are of order 1; their exact values 
depend on the spectrum of the MSSM and have been computed, {\it e.~g.}, in 
Ref.~\cite{esp}.  For new constraints to arise from the requirement of
symmetric vacuum stability at $T \ge T_c$, the inequality 
$T_c \sim 100\ {\rm GeV} < m_t/\sqrt{c_{_L,_R}}$ must be satisfied.  

We note that the high-temperature expansion \cite{dj} is not expected to be
accurate for $T \sim T_c$.  
Therefore, the accuracy to which one can determine the boundaries of the
shaded regions in Fig. \ref{fig1} is limited by one's inability to
determine the effective potential accurately for $T\sim T_c$ due to the 
limitations of the theoretical framework \cite{dj}.
It is clear, however, that the size
of the shaded regions in Fig. \ref{fig1} is rather small, so their
exclusion cannot cause an appreciable reduction in the MSSM parameter space.  
For the generic values of  parameters outside the shaded regions, the
second-order  phase transition into the CCB minimum does not take place.

We have demonstrated that the symmetric vacuum of the MSSM is generally 
stable with respect to second-order phase transitions to a CCB minimum 
at $T>T_c$.  However, the question of a first-order transition at finite
temperature remains open.  The probability of tunneling in the
high-temperature limit is suppressed by the factor \cite{linde}

\begin{equation}
\Gamma/{\sf V} \propto e^{-S_3[\tilde{\phi}]/\hbar T} 
\end{equation}
where $S_3[\tilde{\phi}]$ is the three-dimensional action of the $d=3$ bounce. 
The time allowed for the transition is roughly 
$t_{_T}=m_{_{Pl}}/T^2$, the age of the Universe when the temperature equals
$T$, which means that $S_3[\tilde{\phi}]/\hbar T$ must be less than about
45 for the transition to take place.  

Thus, it appears most likely that if $T_c$ is not too small (of order
$100$ GeV, as is generally believed), then the second-order transition to
the CCB minimum of the type discussed above is not likely, and the
Universe is driven towards the color and charge conserving SML
minimum of the scalar potential.  While, admittedly, this is not a rigorous
theorem, the color and charge conserving minimum appears to be favored by
the thermal evolution of the Universe.  We leave the detailed investigation
of this issue for future work. 

Our next question is what happens after the Universe cools down in the 
SML minimum of the potential: 
will the false vacuum be effectively stable at $T\approx 0$ even in
the presence of some deeper CCB minima, or not.

\section{Numerical analysis of tunneling rates at zero temperature}

To make a definitive determination of whether the SML vacuum is
stable with respect to decay into a lower CCB minimum, one has to compute
the tunneling probability numerically.  Analytic computation is usually
feasible only in the thin wall limit.   

In the semi-classical limit, the zero-temperature tunneling probability 
\cite{vko,c,cc,ak2,klw} per unit volume is 

\begin{equation}
\Gamma/{\sf V}= C_{_G} \left [\int |\bar{\phi}(x)|^2 d^4x \right ]^{N/2}
\left(\frac{S[\bar{\phi}]}{2\pi \hbar} \right)^2 
e^{-S[\bar{\phi}]/\hbar} 
\left |
\frac{\det'[-\partial_\mu^2+U''(\bar{\phi})]}{\det[-\partial_\mu^2+U''(0)]}
\right |^{-1/2}
\times (1+O(\hbar))
\label{one}
\end{equation}
where $S[\bar{\phi}]$ is the Euclidean action of the bounce, a solution to
the variational equation $\delta S=0$, $\det'$ stands for the determinant
with all the zero eigenvalues omitted, $N$ is the number of Goldstone zero 
modes and the $C_{_G}$ is the group-theoretical coefficient \cite{ak2,klw}. 

Suppose the scalar potential $U(\phi_1, ... , \phi_n)$ has a local minimum
at $\phi_i=\phi_i^f, \ i=1,2,...,n$,  as well as at least one
additional (local or global) minimum at $\phi_i=\phi_i^t, \ i=1,2,...,n$;
$U(\phi^t)<U(\phi^f)$.  Then the bounce
$\bar{\phi}(x)=(\bar{\phi}_1(x), ...,\bar{\phi}_n(x) )$ is a non-trivial
$O(4)$-symmetric \cite{c_comm} solution $\bar{\phi}(r),\ r=\sqrt{x^2}$, of
the system of Euler-Lagrange equations: 

\begin{equation}
   \Delta \bar{\phi}_i(r)= \frac{\partial}{\partial \bar{\phi}_i}
U(\bar{\phi}_1, ... , \bar{\phi}_n) 
\label{bounce_eq}
\end{equation}
with the following boundary conditions:

\begin{equation}
\left \{ \begin{array}{l}
    (d/dr) \bar{\phi}_i(r)|_{r=0}=0 \\  \\
   \bar{\phi}_i(\infty)=\phi^f
        \end{array} \right.
\label{boundary}
\end{equation}

In the case of a potential that depends on a single scalar field, one can
solve equation (\ref{bounce_eq}) with the boundary condition  
(\ref{boundary}) numerically.  The straightforward technique is to
assume some value for the unknown quantity $\bar{\phi}(0)=\phi^e$, the so
called  ``escape point''.  Then one can integrate equation (\ref{bounce_eq})
numerically and vary the escape point until the proper limit
$\bar{\phi}_i(\infty)=\phi^f$ is reached.  Here one uses the fact that if 
for some value $\phi^e=\phi^e_-$ the corresponding $\bar{\phi}_i(\infty)> 
\bar{\phi}^f$, and for some other value $\phi^e=\phi^e_+$,
$\bar{\phi}_i(\infty)< \bar{\phi}^f$, then the true escape point must lie
somewhere between $\phi^e_-$ 
and $\phi^e_+$.  This strategy fails when one has to deal with a potential
that depends on several scalar fields.  The peculiarity of 
one-dimensional topology no longer allows one to find the true escape
point as the compromise between the ``undershoot'' and the ``overshoot''
\footnote{Even in the case of a single scalar field, the shooting method
may turn out to be ineffective for a potential with sufficiently degenerate
minima.  This is because one may be required to specify the trial value for
the escape point with an exceedingly high precision to compute the action
of the bounce to a given accuracy.
}. 
It is a generic property of the bounce that it is a saddle point of the
Euclidean action \cite{c}, and not a minimum, and thus small changes in
the initial conditions result in large changes in the form of the solution.
Therefore, it is impossible to find the bounce numerically in the case of a
multi-component field using the procedure just mentioned.

This difficulty was realized by the authors of \cite{chh}, who proposed an
iterative procedure to look for the bounce as a special point of the
discretized action on a lattice.  Since the solution in question is not a
minimum, but a saddle point, they tried to minimize the action with respect
to random variations, while maximizing the same action
with respect to scaling $r \rightarrow \lambda r$.  We find that the
iterative procedure of this kind is ill-defined and cannot have a meaningful
limit.  First, it is impossible to separate the variations
corresponding to scaling from those orthogonal to scaling in a practical
numerical simulation.  Second, although it is true 
that the bounce maximizes the action with respect to scaling, it is easy to
see that the generator of such a variation cannot be the eigenvector of the
second variation operator corresponding to the negative eigenvalue.  

Instead, we use a new method proposed in \cite{ak1} to find the solution of
(\ref{bounce_eq}),(\ref{boundary}).  The idea is to 
turn the saddle point of a Euclidean action into a true minimum by
adding to it some auxiliary terms which vanish for $\phi=\bar{\phi}(x)$.
The resulting ``improved action'' will have a minimum at the point
corresponding to the desired solution, which can now be found by minimizing
the discretized version of improved action on the lattice. 
Details of the application of this method are given in Appendix A.

\section{The MSSM vacuum stability at zero temperature}

For the reasons explained above, we consider the MSSM potential with the
third generation only.  We also neglect all of the trilinear couplings
except those proportional to the largest Yukawa couplings, $y_t$ and $y_b$.  
This justifies dropping the squarks of the first two generations.
Also, although we allow the values of ${\rm tan}\beta$ from $\sim 1$ to 
$\sim 60$, the most stringent constraints come from the small ${\rm
tan} \beta $ region, where the top Yukawa coupling is larger.  As was
pointed out in Ref. \cite{clm}, the noteworthy CCB minima can only develop
in the directions along which one and only one of the trilinear terms is
non-zero.  Thus, it is the size of the largest Yukawa coupling that 
affects the lifetime of the false vacuum the most.  
We therefore consider the following subset of the MSSM scalars:
$\Phi=\{H_1^0, H_2^0, H_1^-, H_2^+, \tilde{Q}_{_L}, \tilde{t}_{_R},
\tilde{b}_{_R} \}$, where 
$\tilde{Q}_{_L}=\{ \tilde{t}_{_L},\tilde{b}_{_L} \}$. 

The superpotential

\begin{equation}
W=y_t Q_{_L} H_2 t_{_R} +y_b Q_{_L} H_1 b_{_R} - \mu H_1 H_2
\label{sptnb}
\end{equation}
corresponds to the following scalar potential (with the soft SUSY breaking
terms included) at tree level:

\begin{equation}
V=V_{2,H}+V_{2}+V_3+V_4,
\label{Vb}
\end{equation}
where $V_{2,H}$ comprises the terms quadratic in the Higgs fields, 

\begin{eqnarray}
V_2 & = &
+ m_{\tilde{t}_{_L}}^2 \tilde{t}_{_L}^2 + m_{\tilde{t}_{_R}}^2
\tilde{t}_{_R}^2 + m_{\tilde{b}_{_L}}^2 \tilde{b}_{_L}^2 
+ m_{\tilde{b}_{_R}}^2 \tilde{b}_{_R}^2, \label{V2b} \\
V_3 & = & - y_t A_t (H_2^0 \tilde{t}_{_L} -H_2^+\tilde{b}_{_L}) \tilde{t}_{_R} 
- y_b A_b (H_1^- \tilde{t}_{_L} - H_1^{0} \tilde{b}_{_L} )  \tilde{b}_{_R}
\nonumber \\
& & -  y_t \mu (H_1^0)^* \tilde{t}_{_L} \tilde{t}_{_R} -  y_b \mu (H_2^+)^*
\tilde{t}_{_L} \tilde{b}_{_R} 
\nonumber \\ & & 
-  y_t \mu (H_1^-)^* \tilde{b}_{_L} \tilde{t}_{_R} -  y_b \mu (H_2^0)^*
\tilde{b}_{_L} \tilde{b}_{_R} +  
{\rm h.\ c.},
\label{V3b}  
\end{eqnarray}
and $V_4$ is comprised of the D-terms, as well as the terms of the form 
$y_{t,b}^2 \phi_1^2 \phi_2^2 $, where $\phi_1 \in \Phi$. 

We now come to the issue of radiative corrections.  It has been argued
(see, {\it e.\,g.}, \cite{b}) that if the CCB minimum occurs at a scale
which is within one or two orders of magnitude from the electroweak scale,
then the (logarithmic) radiative corrections are too small to cause a
noticeable distortion in the shape of the effective potential, and
therefore can be ignored.  While this conclusion is correct, we would like
to emphasize that, as was discussed in Section 2, it is not so much
the position of the CCB minimum, as the size of the bounce and the value of
the escape point that determine the scale to which the bounce will
``probe'' the effective potential.  The shape of the effective potential
beyond that scale has no effect on the tunneling rate.  In our numerical 
calculations, a typical size of the bounce for realistic values of of the
MSSM parameters is of order $(1-0.1\  {\rm TeV})^{-1}$ (except for
deliberately fine-tuned cases of highly degenerate minima in which the
bounce blows up to the size of its thin-wall limit, $R\sim 1/\epsilon$, and
the tunneling becomes highly improbable).  This justifies {\it post factum}
the neglect of the one-loop contribution in the effective potential. 
As long as one never encounters very small (in length units) bounces, one
can be certain that only the shape of the potential around the electroweak 
scale is relevant for tunneling.

\begin{figure}
\postscript{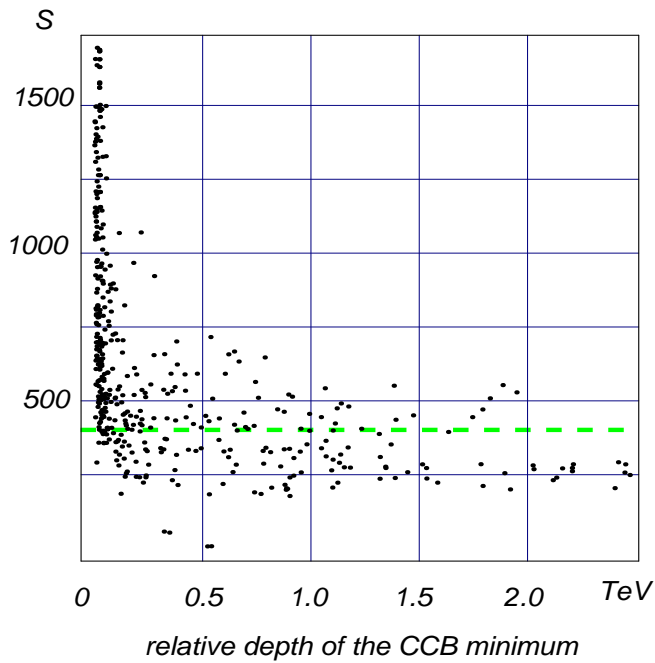}{0.4}
\caption{The action of the bounce, {$S$}, does not depend sensitively on the
depth of the CCB minimum, {$\Delta V$}, except in the ``thin-wall''
limit (small  {$\Delta V$}).  
}
\label{svfig}
\end{figure}

We search the MSSM parameter space by generating randomly the values  of
the parameters that enter in the scalar potential (\ref{Vb}). 
Then the minima are found numerically and the tunneling probability is
computed using the method of Appendix A.

The results of the numerical analyses are plotted in Figure \ref{svfig},
where the action of the bounce $S$ is shown as a function of the relative
depth of the CCB minimum with respect to the SML minimum, while the other
MSSM parameters are varied randomly.  For each point plotted, the global
minimum of the potential is not color and charge conserving.  Nevertheless, 
one observes that $S$ takes values on both sides of the critical value,
$S=400$.  Therefore,  for some values of parameters, namely those for which
$S>400$,  the false SML vacuum is stable on a time scale large compared
to the age of the Universe.  Such color and charge breaking minima are
``safe'', and the corresponding parameters are allowed.  In contrast, those
points that correspond to $S<400$ are ruled out by the mere existence of
the world as we know it.  Another lesson one learns from Figure \ref{svfig}
is that the tunneling rate is not very sensitive to the depth of the CCB
minimum, except in the limit of nearly degenerate minima, the so called 
``thin-wall'' limit.  This is to be expected, as was discussed in 
Section 2.

\begin{figure}
\postscriptfx{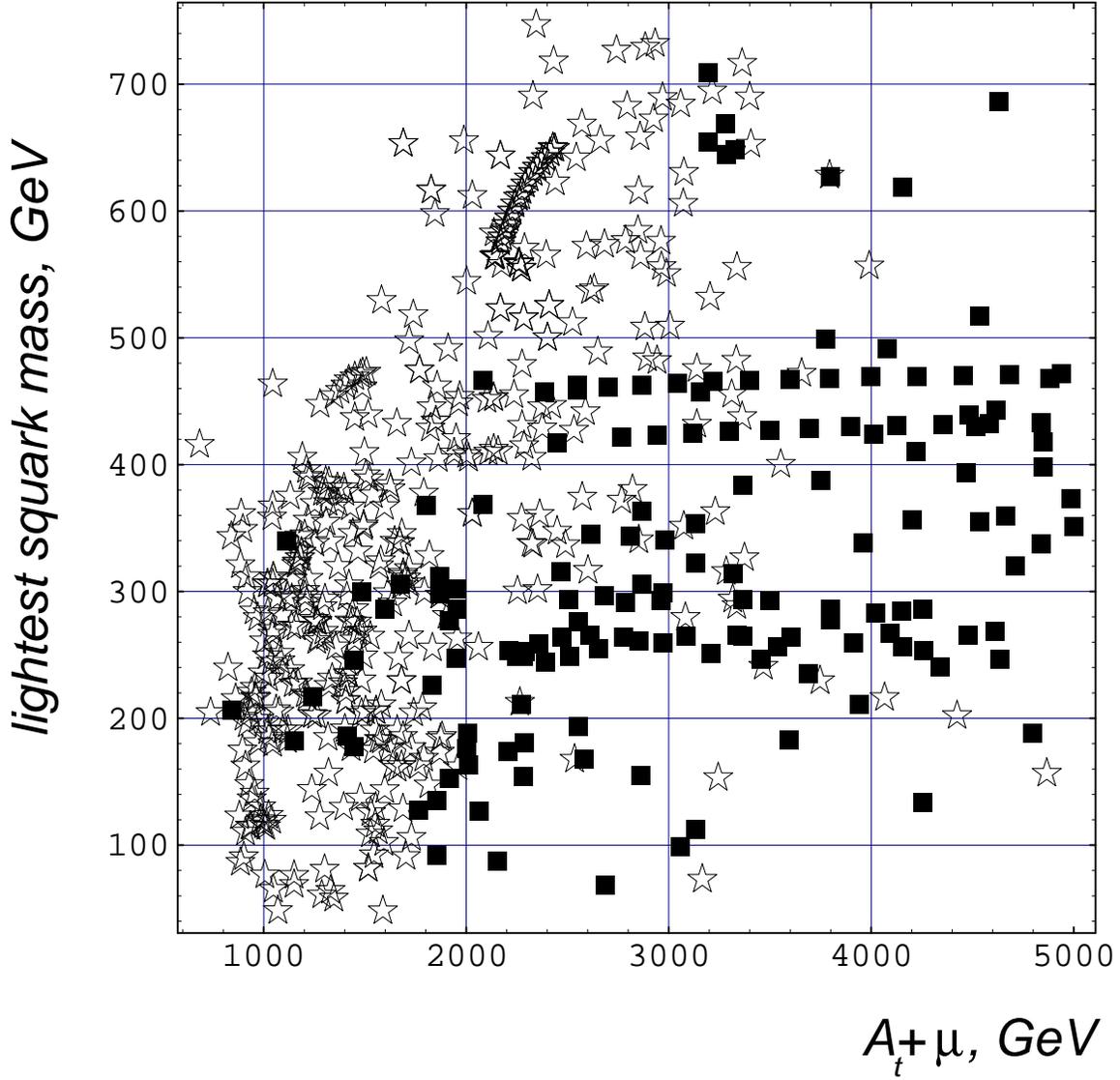}
\caption{The domains of stability (stars) and instability (boxes) of the
false SML vacuum with respect to tunneling into the global CCB minimum. 
Light top squark and large trilinear couplings generally correspond to a
lower and thinner barrier and, thus, higher probability of tunneling. 
}
\label{amusqfig}
\end{figure}

The domain of stability of the false SML vacuum with respect to tunneling 
is delineated by stars in Figure \ref{amusqfig}.  The lighter top squark 
in the presence of the large trilinear couplings forces the barrier, which 
keeps the system in the metastable vacuum, to be thinner and lower. 
That results in a higher likelihood of tunneling.  The points labeled by
boxes fall into the domain that is excluded by the existence of our
(color and charge conserving) Universe.  On the other hand, if both 
left-handed and right-handed stops are heavy, and if the trilinear terms
are small, then the false vacuum is stable and the presence of a global CCB
minimum is irrelevant.  We note that a number of models favor the left
lower corner of the plot in Figure \ref{amusqfig}, where only a few
``dangerous'' CCB minima occur. 

\begin{figure}
\postscriptfx{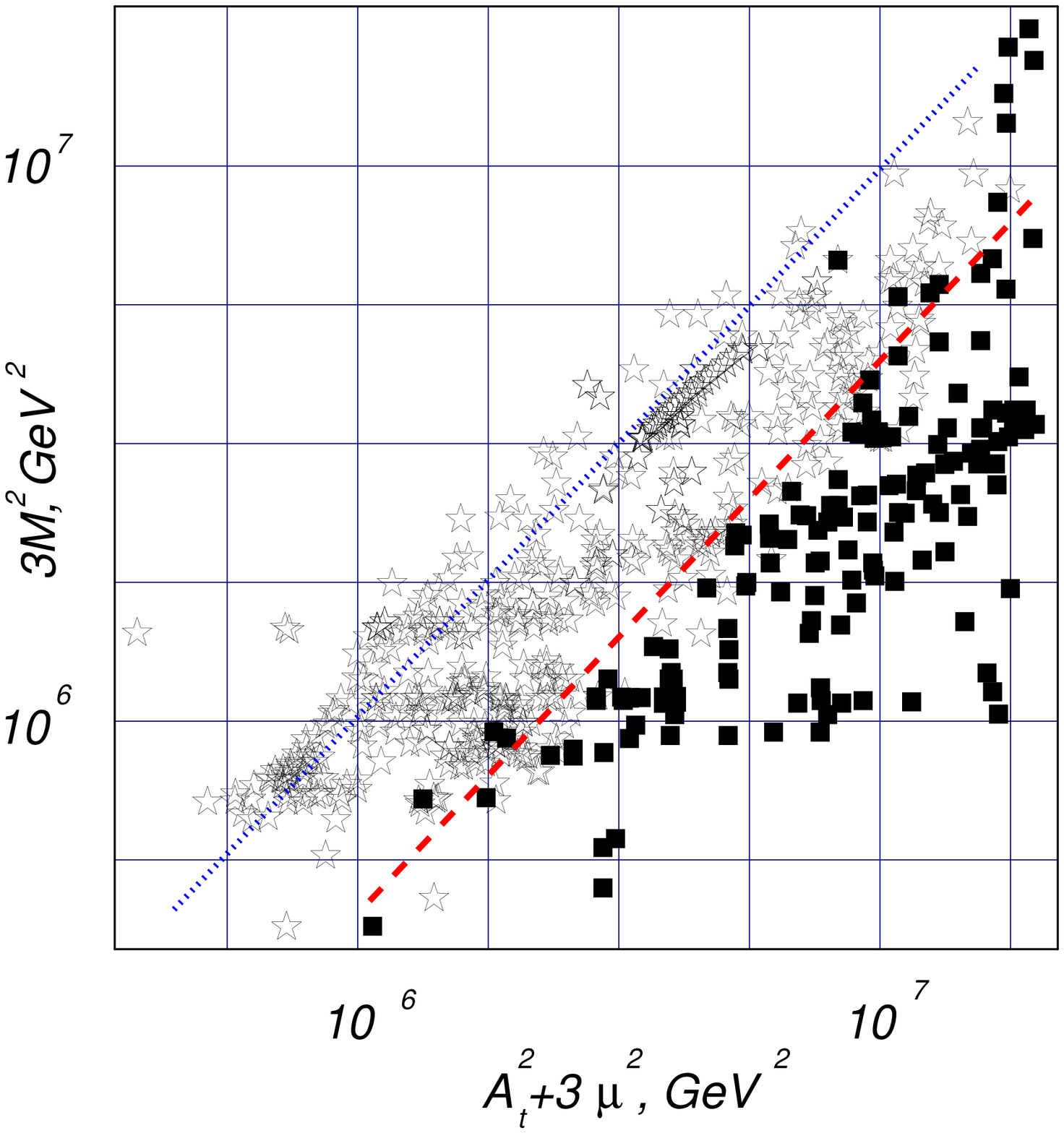}
\caption{ Each point represents the set of the MSSM parameters for which
the global minimum of the potential is color and charge breaking.
The stars correspond to the SML false vacua whose lifetime is large
compared to the age of the Universe.  The boxes indicate those points in
the parameter space for which the false SML vacuum should have decayed via 
quantum tunneling. The dotted line represents the empirical
criterion for the absence of the 
global CCB minima: {$A_t^2+3 \mu^2 < 3 M^2$}, where 
{$ M^2=m_{\tilde{t}_{_L}}^2+m_{\tilde{t}_{_R}}^2$}.  Taking into account
the tunneling rates 
relaxes this constraint to, roughly, {$A_t^2+3 \mu^2 < 7.5 M^2$}, shown as 
the dashed line.  The scale is logarithmic.} 
\label{linefig}
\end{figure}

It would be useful to derive some empirical algebraic constraints to
distinguish between the allowed and excluded domains of parameter space 
based on the numerical results.  For instance, it has been argued 
\cite{clm} that the inequality 

\begin{equation}
A_t^2+3 \mu^2 < 3 (m_{\tilde{t}_{_L}}^2+m_{\tilde{t}_{_R}}^2)
\label{neq}
\end{equation}
is a reasonably good condition for excluding a global CCB minimum.  
This inequality would force one to be above the dotted line in 
Figure~\ref{linefig}.  This is in good agreement with
our numerical results.  However, if one takes into account the tunneling
rates, the constraint (\ref{neq}) is relaxed significantly.  The empirical 
inequality which should replace (\ref{neq}), 

\begin{equation}
A_t^2+3 \mu^2 < 7.5 \: (m_{\tilde{t}_{_L}}^2+m_{\tilde{t}_{_R}}^2),
\label{newneq}
\end{equation}
is depicted by a thick dashed line in Figure~\ref{linefig}.  The simple
condition (\ref{newneq}),  should be applied with caution because, strictly
speaking, it is neither necessary nor sufficient.  It is an 
approximate empirical inequality that may be useful for a crude
determination of whether the CCB minima are ``dangerous''.   

\begin{figure}
\postscript{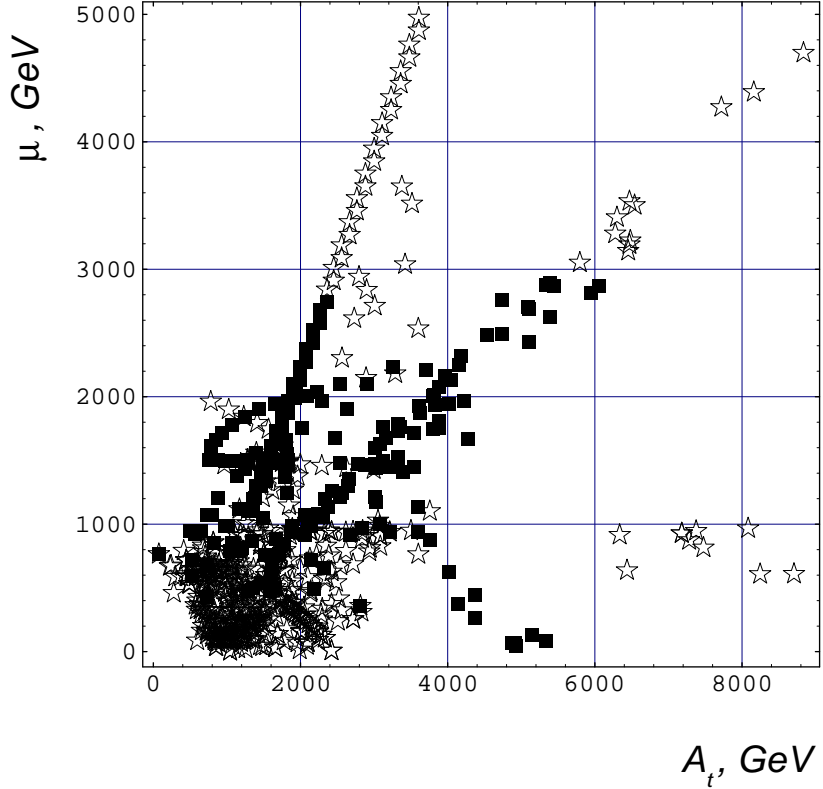}{0.5}
\caption{Tunneling probability for unphysically large values of 
{$A_t$} and {$\mu$}.  As the CCB minimum moves farther away, it becomes 
``less dangerous''. As before, the stars mark the points with {$S>400$}, 
while the boxes depict those with {$S<400$}.  }
\label{amufig}
\end{figure}

For the phenomenologically attractive values of $|\mu|<2$ TeV, 
$|A|<4$ TeV, it is generally true that ``the larger the trilinear coupling, 
the more dangerous is the corresponding CCB minimum''.  However, it is
instructive to examine what happens to the tunneling probability in the
limit of very large $\mu$ and $A_t$ (and large enough squark mass terms to
ensure the existence of the SML minimum).  In that limit, as the CCB
minimum moves away from the SML minimum, the barrier separating the two 
becomes thicker, and the false vacuum should become more stable.  This is,
in fact, what happens.  The set of point in Figure \ref{amufig} includes
those points (located in the lower left corner) shown in Figures
\ref{svfig}, \ref{amusqfig} and \ref{linefig}. In addition, Figure
\ref{amufig} displays the points corresponding to some very large values of
$A_t$ and $\mu$.  As expected, the tunneling probability diminishes for
very large values of $A_t$ and $\mu$, and $m_{\tilde{t}_{_L}}$ and
$m_{\tilde{t}_{_R}}$.   

To summarize, if the global CCB minimum is nearly degenerate with the local
SML minimum (thin-wall limit), then the tunneling probability is extremely
small.  As the trilinear couplings increase, the false vacuum decay rate
increases because the escape point of the bounce moves out of the flat
vicinity of the global minimum into the region in which the gradient of the
potential is significant.  However, a further increase in the size of the
trilinear couplings, as well as the consequent increase in the squark mass 
terms, makes the barrier thicker and pushes the
escape point away from the SML minimum.  This eventually causes a decrease
in the tunneling rate.  In accordance with one's intuition, the low-energy
physics is unaffected by the physics at the very high energy scales.

\section{Conclusion}

The color and charge conserving minimum may not be the global minimum of the
MSSM potential.  It is possible that the Universe rests in a false
vacuum whose lifetime is large in comparison to the present age of the
Universe.  Under fairly general conditions, the SML vacuum may be favored by
the thermal evolution of the Universe, even if it does not represent the
global minimum. 

The existence of the CCB minima of the scalar potential results in some
important constraints on models with low-energy sypersymmetry. 
However, the commonly imposed (see, {\it e.\,g.}, Refs. \cite{cr,clm2})
requirement that the SML minimum be global is too strong and may 
overconstrain the theory.  In fact, for a large portion of the
parameter space the presence of the global CCB minimum is irrelevant
because the time required for the Universe to relax to its lowest energy
state may exceed its present age.  The basic reason for this is that the
quantum  tunneling is a non-perturbative phenomenon that is naturally
associated with the energy scale that is exponentially smaller (suppressed
by a factor $\exp \{-S \}$) than the typical scale in the theory.   

We have computed numerically the SML false vacuum decay rates for a variety
of values of the MSSM parameters.  Our results indicate clearly that
the MSSM vacuum stability with respect to tunneling into a CCB minimum or
a ``UFB direction'', imposes important constraints on models with 
low-energy supersymmetry.  

\section*{Acknowledgments}

We would like to thank M.~Carena, S.~Coleman, M.~Cveti\v{c}, 
J.~R.~Espinosa, R.~Hempfling,
L.~McLerran, H.~Murayama, N.~Polonsky, M.~Shaposhnikov, M.~Shifman, R.~Shrock,
P.~Steinhardt, M.~Voloshin, C.~Wagner and T.~Yanagida 
 for helpful discussions.  
This work was supported by the U.~S.~Department of Energy Contract 
No.~DE-AC02-76-ERO-3071 as well as by the National Science Foundation Grant 
No.~PHY94-07194.   

\appendix
\section*{Appendix A}

The numerical algorithm for computing the tunneling probability
comprises several steps. First, the positions of the minima,
$\phi^t$ and $\phi^f$ of the potential $U(\phi)$ are determined numerically
for a given set of values  of the MSSM parameters.  If there is no minimum
below the SML vacuum, a new set of parameters is chosen.  We define the
Euclidean action on the $L$-point lattice:

\begin{eqnarray}
S[\phi] & = & T[\phi]+V[\phi] \\
T[\phi] & = & 2\pi^2 \Delta^4 \sum_{m=1}^{L-1} (d_{m+1}-d_m) d_m^3
\left
(\sum_{i=1}^n \frac{(\phi_{i}^{(m+1)}-\phi_{i}^{(m)})^2}{2(d_{m+1}-d_m)^2 
\Delta^2} \right ) \\
V[\phi] & = & 2\pi^2 \Delta^4 \sum_{m=1}^{L-1} (d_{m+1}-d_m) d_m^3
U(\phi_1^{(m)},...,\phi_n^{(m)}) 
\end{eqnarray}
where $d_m$ is the position of the $m$'s site of the lattice in
dimensionless units, and $\Delta$ is the length parameter that determines the
overall scale and is chosen so as to optimize the computation.  

The next step is to define the improved action \cite{ak1}, for which the 
bounce is the minimum.  We do that by adding to the action some auxiliary
terms which (i) vanish as $\phi(x)$ approaches the bounce, and (ii) 
make the bounce $\bar{\phi}(r)$ a minimum of the improved action \cite{ak1}:

\begin{equation}
\tilde{S}[\phi] \equiv S[\phi] + \lambda \: \left| T[\phi]+2 V[\phi] \right
|^{1/2} \label{impr_action}
\end{equation}
where $\lambda$ is some arbitrary (dimensionless) Lagrange multiplier. 

\begin{figure}
 {\setlength{\epsfxsize}{6in}
\setlength{\epsfysize}{3.5in}
  \centerline{\epsfbox{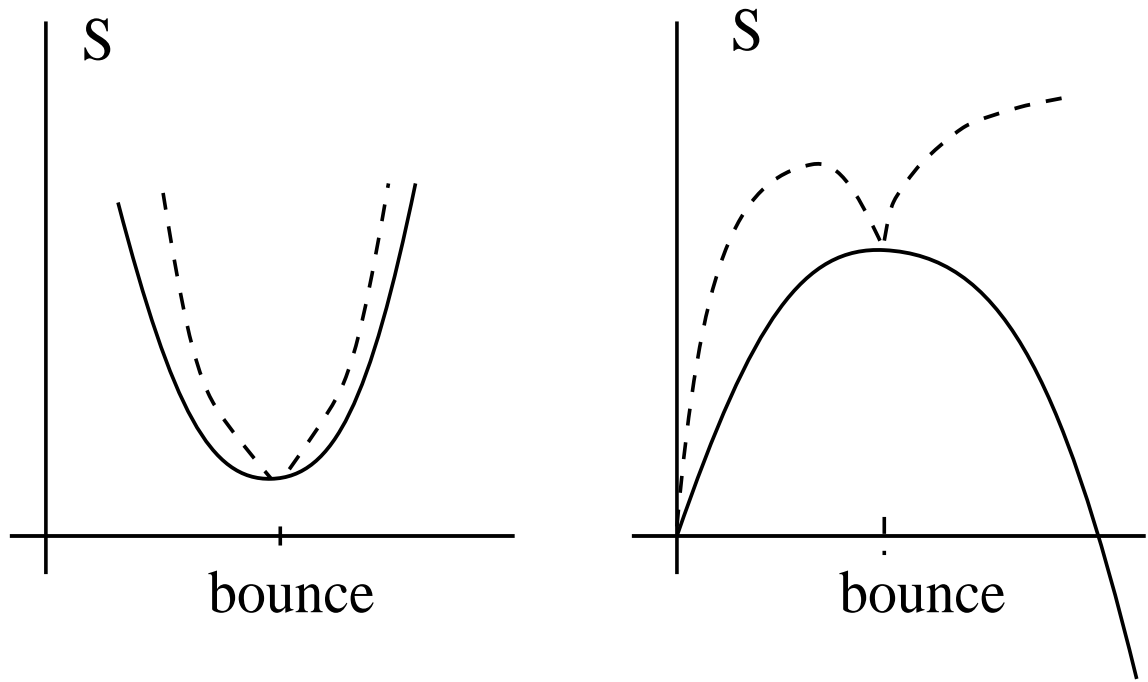}}}
\caption{The improved action {$\tilde{S}[\phi]$} (dashed curve) has a minimum
that coincides with the saddle point of the actual Euclidean 
action {$S[\phi]$} (solid curve), as shown in two projections.  The origin
on the right corresponds to the trivial solution {$\phi(x) \equiv \phi^f$}
for which {$\tilde{S}[\phi]= S[\phi]=0$}. The size of the ``gap'' between
the two curves is a function of $\lambda$. }
\label{ia}
\end{figure}

The effect of adding the auxiliary terms is shown qualitatively in
Fig. \ref{ia}, where two projections of the saddle point of $S[\phi]$
are depicted symbolically. 

The bounce $\bar{\phi}(x)$ is always a minimum of the improved action
$\tilde{S}[\phi]$.  However, unless $\lambda$ is chosen to be large enough, 
the difference $\tilde{S}[\phi]-S[\phi]$ may appear to be too small in the
vicinity of the bounce.  In this case, it is possible that the small
perturbations will take one over the barrier (Fig. \ref{ia}) towards
the trivial solution of zero action $\phi(x) \equiv \phi^f$.  
Thus, it is crucial to take $\lambda$ large enough for the numerical
simulation to succeed.  

Initially, one may choose

\begin{equation}
\phi_{i}^{(m)}=\left \{\begin{array}{l}
                             \phi_{i}^0, \ \ m<L/2 \\
                             \phi_{i}^f, \ \ m \ge L/2, 
                       \end{array} \right.
\end{equation}
where $\phi^0$ is a zero of $V(\phi)$.  Alternatively, one may start from a
different profile for the bounce, {\it e.\,g.} the thick-wall ansatz
described below in the Appendix B. 

Then we find the optimized value for the overall scale $\Delta$ by
maximizing the action with respect to $\Delta$.  After that, we allow random
variations of the values of the scalar field at each lattice site to
minimize the improved action (\ref{impr_action}).  
The iterations stop when further
variations do not lead to a reduction in the improved action.  As an 
independent criterion of the quality of the fit, we require that 
$T[\phi]/(-2 \: V[\phi])=1 \pm 0.01$ (see, {\it e.\,g.}, Ref. \cite{ak1} and
references therein for a discussion of this identity). 

As was explained in Sections 2 and 5, tunneling into a ``UFB direction'' 
is equivalent to a transition into a very deep CCB 
minimum.  The details of the effective potential at scales much larger
than $\phi^e$ do not affect the decay probability.  In particular, it
is irrelevant whether the alleged ``UFB direction'' is heading towards a
very deep minimum, or minus infinity.  This allows one to introduce an
effective cutoff to stop the run-away fields from going to infinity.  In
practice, we did not allowed the value of $\phi$ to run beyond $10$ TeV. 
Therefore, the ``UFB directions'' were treated as if they lead to a CCB
minimum with a $10$ TeV vev.  In each particular case this procedure is
justified {\it post factum} by ensuring that the value of the escape point 
is small compared to a $10$ TeV cutoff.

\appendix
\section*{Appendix B}

In the limit of nearly degenerate minima separated by a high barrier, the
so-called thin-wall limit, the bounce can be approximated by a smoothed out
step function \cite{c}.  This approximation proved very useful in
estimating the tunneling rates and was used in our analysis of the toy
model in Section 2.  In practice, however, one rarely encounters a situation
in which the thin-wall approximation is in good agreement with the numerical
results ({\it c.\,f.} Ref. \cite{chh}).  The tunneling rates
are usually very small in the thin-wall limit, and therefore in many
physically interesting models the first-order phase transition takes place 
when the energy difference between the true and the false vacuum is not
small in comparison to the height of the barrier.  Then it is necessary to
go beyond the thin-wall approximation, which is usually done by means of
a numerical calculation. 

If, however, the energy difference $\Delta V$ between the two vacua is
larger than the height of the barrier, one can find a simple approximation
to the bounce in what we call a ``thick-wall'' limit.  As  $\Delta V$
increases, the so-called ``escape point'', the value of the field in the
center of the bounce $\phi_e \equiv \bar{\phi}(0)$, moves away from the
minimum of the potential.  Then equation (\ref{bounce_eq}) can be
linearized in the vicinity of the escape point because $\partial
U(\phi)/\partial \phi$ is a constant 
independent of $\phi$ for $\phi \approx \phi(0)$.  Therefore, in the
vicinity of the center of the $O(4)$-symmetric bounce $\bar{\phi}(r)$, 

\begin{equation}
\left \{ 
\begin{array}{l}
\bar{\phi}''(r)+\frac{3}{r} \bar{\phi}'(r)=-a \\ \\
\bar{\phi}'(0)=0, 
\end{array} \right.
\label{lin}
\end{equation}
where the constant $a=\left | \frac{\partial U}{\partial \phi} (\phi_e)
\right |$.  This is a linear equation whose only solution satisfying the
boundary condition is of the form

\begin{equation}
\bar{\phi}(r)=-\frac{a}{8} r^2+ \phi_e
\label{zero}
\end{equation}

Outside the small neighborhood of the origin, the approximation (\ref{lin})
is not valid and $\bar{\phi}(r)$ falls off exponentially, just as in the
thin-wall limit \cite{c}.  The equation for the bounce $\bar{\phi}(r)$ for
large $r$ (and, therefore, small $\phi$) becomes

\begin{equation}
\left \{ 
\begin{array}{l}
\bar{\phi}''(r) = \partial U(\phi)/\partial \phi \approx m^2 \bar{\phi}^2
\\ \\ 
\bar{\phi}(\infty)=0, 
\end{array} \right.
\label{large_r}
\end{equation}
where $m$ is the mass term for the potential $U(\phi)=(m^2/2) \phi^2+...$. 
The solution of the equation (\ref{large_r}) is 

\begin{equation}
\bar{\phi}(r)=C e^{-mr},
\label{large}
\end{equation}
where $C$ is an arbitrary constant.  We now sew the approximate solution
for the bounce from the two asymptotics, (\ref{zero}) and (\ref{large}), at
some point $r=R$.  The values of $C$ and $R$ are determined by requiring
continuity and differentiability of the solution at $r=R$.  The resulting 
{\it ansatz} is 

\begin{equation}
\bar{\phi}(r) = \left \{ 
\begin{array}{ll}
       \phi_e-(a/8) r^2, & r<R \\ \\
        \frac{a R}{4m} e^{-m(r-R)}, & r \ge R, 
\end{array} \right.
\label{ansatz}
\end{equation}
where 

\begin{equation}
R=\frac{1}{m} [\sqrt{1+8m^2 \phi_e/a} -1],
\end{equation}
and the value of $\phi_e$ is an unknown parameter which can be found either
from requiring that 

\begin{equation}
0= \frac{d}{d \phi_e} S =\frac{d}{d \phi_e} (T+V),
\end{equation}
or, equivalently, from solving the equation

\begin{equation}
T=-2 V,
\end{equation}
where

\begin{equation}
T=2 \pi^2 \int_{0}^{\infty} r^3 dr \ \frac{1}{2} \left (
\frac{d \bar{\phi}(r)}{dr} \right )^2 \ {\rm and } \ V=2 \pi^2
\int_{0}^{\infty} 
r^3 dr \ U(\bar{\phi}(r))  
\label{T_V_thick}
\end{equation}
are functions of $\phi_e$. 

The representation of the bounce in the thick-wall limit described above is
approximate and is not very useful in application to the MSSM. 
However, it exhibits the essential features of tunneling in this limiting
case.  We also found it convenient to use this approximate solution as an 
initial profile for the bounce in the numerical procedure described in
Appendix A.

\end{document}